# Ki-67 Index Measurement in Breast Cancer Using Digital Image Analysis

Hsiang-Wei Huang[1]　　Wen-Tsung Huang[2]　　Hsun-Heng Tsai[3*]

1 Department of Photonics, National Chiao Tung University  
2 Division of Hematology and Oncology, Liouying Chi Mei Hospital  
3 Department of Biomechatronics, National Pingtung University of Science and Technology  
*Email: heng@mail.npust.edu.tw

## Abstract

Ki-67 is a nuclear protein that can be produced during cell proliferation. The Ki-67 index is a valuable prognostic variable in several kinds of cancer. In breast cancer, the index is even routinely checked in many patients. Currently, pathologists use the immunohistochemistry method to calculate the percentage of Ki-67 positive malignant cells as Ki-67 index. The higher score usually means more aggressive tumor behavior. In clinical practice, the measurement of Ki-67 index relies on visual identifying method and manual counting. However, visual and manual assessment method is time-consuming and leads to poor reproducibility because of different scoring standards or limited tumor area under assessment. Here, we use digital image processing technics including image binarization and image morphological operations to create a digital image analysis method to interpretate Ki-67 index. Then, 10 breast cancer specimens are used as validation with high accuracy (correlation efficiency r = 0.95127).

With the assistance of digital image analysis, pathologists can interpretate the Ki-67 index more efficiently, precisely with excellent reproducibility.

## 1. Introduction

Ki-67 is a kind of protein that is associated with cellular proliferation. It is considered to be one of the most promising but also controversial biomarkers for proliferating tumors due to its variation in visual assessment and its lack of reproducibility. Although the International Ki-67 in Breast Cancer Working Group has published a series of guidelines for the assessment of the Ki-67 index in clinical practice including scoring method and data analysis steps, most of the laboratories still show little reproducibility due to several reasons including different scoring area, deviation in visual assessment method, and variation in the observed fields under microscopes.

Therefore, a standardized method for the Ki-67 index assessment with high accuracy and reproducibility is crucially needed. In this study, we use digital image processing technics including image binarization and image morphological operations to determine the fraction of Ki-67 positive malignant tumor cells. In addition, ten different samples of known Ki-67 index breast cancer specimens were used for analysis and to validate our digital image analysis system.

## 2. Materials and methods

2.1. Tissue Specimens

Ten Ki-67 staining pictures with different Ki-67 index were recruited from the ChiMei Medical Center, Liouying. All of the pictures were taken from breast cancer specimens.

2.2. Morphometric Analysis

In our study, we use the RGB color model and the HSV model for the Ki-67 index assessment. After setting a suitable threshold, we use image binarization to separate the image background and the stained deep color nucleus, obtaining an image with only stained nucleus.

To get better accuracy in tumor cells counting, we use image morphological operations including dilation and erosion to add or remove pixels to the boundaries of cells in the image. Since the Ki-67 index is defined by the fraction of Ki-67-positive tumor cells, we use the following equation to calculate the ratio of positive tumor cells to normal tumor cells:

$$y_1 = \frac{\sum_{i=1}^{N}\sum_{j=1}^{N} x(i,j)}{\sum_{i=1}^{N}\sum_{j=1}^{N}[C(i,j)+x(i,j)]} \tag{1}$$

where $y_1$ = the fraction of Ki-67-positive tumor cells, $x(i,j)$ = area of the red nucleus, $C(i,j)$ = area of the blue nucleus, and $N$ = image element number.

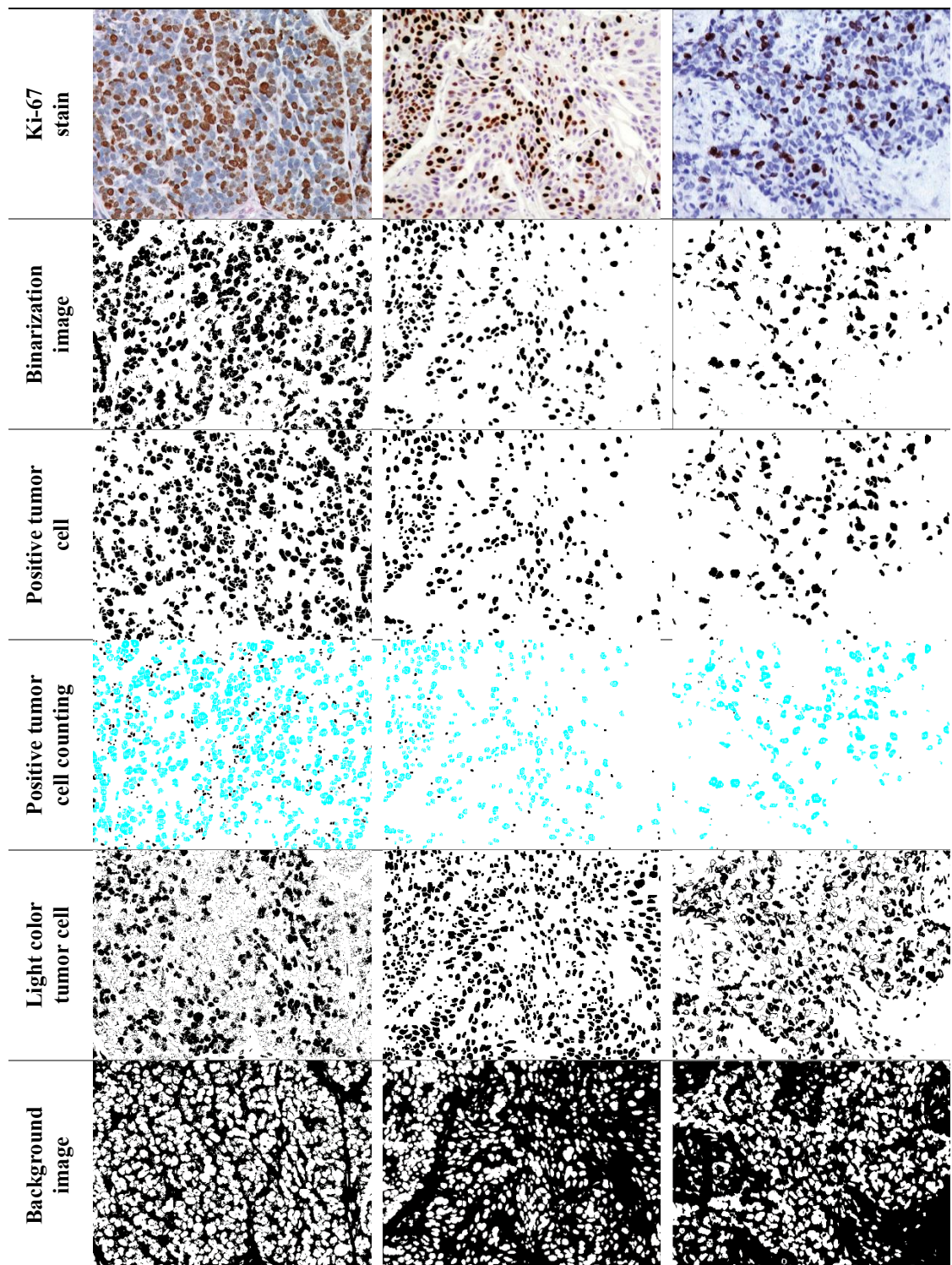

**Figure.1** Image processing of the Ki-67 index computer assessment

We use binarization to separate the background of the original pictures of Ki-67 staining cell and its unstained cells. After binarization, we can get a better image of positive tumor cell image by erosion and dilation. Finally, we use digital counting for the final process of the Ki-67 index assessment.

## 3. Results and Discussion

Figure 1 shows some of the representative images for digital image analysis method. The analysis results of the ten samples are listed in Table 1, including the number of stained cells, unstained cells and Ki-67 index assessment results under different methods including digital image analysis and visual assessment. The median counted number of tumor cells by digital image analysis is 2449, ranging from 1161 to 2882. The Ki-67 index of these 10 samples range from 0.3% to 33.4% according to digital image analysis results.

**Table.1** Summary of Analysis Results in Two Different Methods

| TEST CASE | 1 | 2 | 3 | 4 | 5 | 6 | 7 | 8 | 9 | 10 |
|---|---|---|---|---|---|---|---|---|---|---|
| NUMBER OF STAINED CELL | 77 | 240 | 471 | 442 | 771 | 79 | 189 | 80 | 9 | 282 |
| NUMBER OF UNSTAINED CELL | 1883 | 2473 | 2278 | 2013 | 1534 | 1082 | 2631 | 2363 | 2329 | 2600 |
| DIGITAL ANALYSIS KI-67 INDEX | 3.9% | 8.8% | 17.1% | 18.0% | 33.4% | 6.8% | 6.7% | 3.2% | 0.3% | 9.7% |
| VISUAL ANALYSIS KI-67 INDEX | 5.0% | 7.5% | 25.0% | 24.0% | 40.0% | 7.5% | 7.5% | 5.0% | 3.0% | 7.5% |

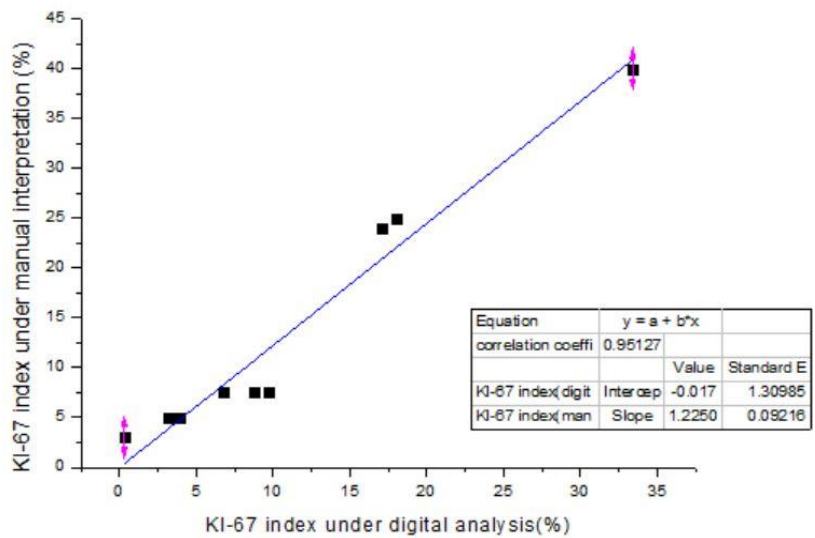

**Figure2.** Scatter Diagram of Analysis Results in Visual Assessment and Digital Image Analysis and its Correlation Analysis

Figure 2 shows the scatter diagram of analysis results in visual assessment and digital image analysis and its correlation analysis. The two different methods in Ki-67 index assessment shows a high correlation (r = 0.95127), indicating digital analysis can produce high accuracy in Ki-67 index assessment. In daily clinical practice, the pathologist usually counts 500 to 1000 tumor cells to determine the percentage of Ki-67 nucleus staining positive cells. In our samples, using digital image analysis method, much more tumor cells can be included for reading in a very short time. Thus, sampling bias and time consumption can be avoided.

## 4. Conclusion

Although Ki-67 has been considered to be an important biomarker in breast cancer therapy because of its ability to provide prognostic and predictive information, it still has limitations because of its lack of reproducibility and stability in visual assessment and time-consuming. In this study, we offer a digital image analysis method with high accuracy and reproducibility. Using this method in clinical practice or scientific research could improve the standardization of the Ki-67 scoring assessment.

There are some limitations in our study. In this study, there are only ten samples available for assessment and all the Ki-67 index of tumor tissue are less than 50%, although the results show high consistency with those visual assessment results by experienced pathologists. Further studies are needed to confirm whether digital analysis can still perform high accuracy and reproducibility in Ki-67 index interpretation under different breast tumor types.

In conclusion, the digital analysis method possesses the comparable results of the visual assessment method in identifying tumor cells, Ki-67 staining tumor cells, the ability of automatic counting, and furthermore high reproducibility. We consider this to be a promising technique and can offer a good alternative method with reliable accuracy, reproducibility and efficiency for pathologists to interpretate Ki-67 index in daily clinical practice.

**Acknowledgements**

This study was supported by Division of Hematology and Oncology, Liouying Chi Mei Hospital.